\documentclass[fleqn,usenatbib]{mnras}

\usepackage{newtxtext,newtxmath}
\usepackage[T1]{fontenc}

\DeclareRobustCommand{\VAN}[3]{#2}
\let\VANthebibliography\thebibliography
\def\thebibliography{\DeclareRobustCommand{\VAN}[3]{##3}\VANthebibliography}

\usepackage{graphicx}	% Including figure files
\usepackage{amsmath}	% Advanced maths commands
\usepackage{wasysym}           
\usepackage{graphicx}
\usepackage{epstopdf}
\usepackage{mathrsfs}
\usepackage{anyfontsize}
\usepackage{natbib}
\usepackage{color}
\usepackage{lipsum}
\usepackage{diagbox}

\title[The tidal disruption radius]{A simple and accurate prescription for the tidal disruption radius of a star and the peak accretion rate in tidal disruption events}

\author[Coughlin \& Nixon]{
Eric R.~Coughlin$^{1}$\thanks{E-mail: ecoughli@syr.edu} and
C.~J.~Nixon$^{2}$
\\
$^{1}$Department of Physics, Syracuse University, Syracuse, NY 13244, USA \\
$^{2}$School of Physics and Astronomy, University of Leicester, University Road, Leicester, LE1 7RH, UK
}

\date{Accepted XXX. Received YYY; in original form ZZZ}

\pubyear{2022}

\begin{document}
\label{firstpage}
\pagerange{\pageref{firstpage}--\pageref{lastpage}}
\maketitle

\begin{abstract}
A star destroyed by a supermassive black hole (SMBH) in a tidal disruption event (TDE) enables the study of SMBHs. We propose that the distance within which a star is completely destroyed by a SMBH, defined $r_{\rm t, c}$, is accurately estimated by equating the SMBH tidal field (including numerical factors) to the maximum gravitational field in the star. We demonstrate that this definition accurately reproduces the critical $\beta_{\rm c} = r_{\rm t}/r_{\rm t, c}$, where $r_{\rm t} = R_{\star}\left(M_{\bullet}/M_{\star}\right)^{1/3}$ is the standard tidal radius with $R_{\star}$ and $M_{\star}$ the stellar radius and mass and $M_{\bullet}$ the SMBH mass, for multiple stellar progenitors at various ages, and can be reasonably approximated by $\beta_{\rm c} \simeq \left[\rho_{\rm c}/(4\rho_{\star})\right]^{1/3}$, where $\rho_{\rm c}$ ($\rho_{\star}$) is the central (average) stellar density. We also calculate the peak fallback rate and time at which the fallback rate peaks, finding excellent agreement with hydrodynamical simulations, and also suggest that the partial disruption radius -- the distance at which any mass is successfully liberated from the star -- is $\beta_{\rm partial} \simeq 4^{-1/3} \simeq 0.6$. For given stellar and SMBH populations, this model yields, e.g., the fraction of partial TDEs, the peak luminosity distribution of TDEs, and the number of directly captured stars. 
\end{abstract}


\section{Introduction}
The tidal disruption of a star by a supermassive black hole (SMBH), known as a tidal disruption event (TDE; e.g., \citealt{rees88, gezari21}), fuels a luminous flare in the center of a galaxy that can offer insight into SMBH properties, stars in galactic nuclei, and accretion physics (including the launching of relativistic outflows; \citealt{giannios11, bloom11, zauderer11, cenko12, brown15}). The detection rate of TDEs is rapidly growing (e.g., \citealt{holoien19, nicholl19, wevers19, hung20, hinkle21, vanvelzen21, hammerstein22}), and is set to explode in the era of the Rubin Observatory \citep{ivezic19}, but the power of a TDE to provide this insight hinges on our ability to reliably interpret observations with theory.

One prediction of TDE theory is that the star is destroyed by tides if it comes within a distance $r_{\rm t}$, the tidal radius, of the SMBH. The timescale for the stellar debris to return to the SMBH -- known as the fallback time -- and the resultant accretion luminosity can then be estimated as (\citealt{lacy82} and Section \ref{sec:analytic} below)

\begin{equation}
T_{\rm fb} \simeq \left(\frac{r_{\rm t}^2}{2R}\right)^{3/2}\frac{2\pi}{\sqrt{GM_{\bullet}}}, \quad L_{\rm fb} \simeq \frac{M_{\star}}{T_{\rm fb}}. \label{Tfb}
\end{equation}
Here $R$ is the characteristic size of the star at the time it reaches the tidal radius, $M_{\bullet}$ is the SMBH mass, and $M_{\star}$ is the mass of the original star. $T_{\rm fb}$ represents the fundamental evolutionary timescale of a TDE, and accurately constraining it therefore amounts to determining the values of $r_{\rm t}$ and $R$. Typically, $r_{\rm t}$ is estimated by equating the tidal force of the SMBH to the surface gravity of the star and dropping numerical factors, which yields 

\begin{equation}
r_{\rm t} = R_{\star}\left(M_{\bullet}/M_{\star}\right)^{1/3}, \label{rtdef}
\end{equation}
and $R = R_{\star}$, where $R_{\star}$ is the stellar radius. Because the tidal force varies as the inverse cube of the distance to the SMBH, Equation \eqref{rtdef} should be correct to within a factor of the order unity, and numerical simulations have confirmed that this is indeed the case over a wide range of stellar type (e.g., \citealt{guillochon13, gafton15, mainetti17, golightly19, lawsmith19, gafton19, lawsmith20, miles20, nixon21}).

However, while the precise distance at which the star is destroyed by tides must be $\sim r_{\rm t}$, the dependence of Equation \eqref{Tfb} on $r_{\rm t}^3$ implies that small changes in $r_{\rm t}$ from its approximate value can have large bearing on the observable properties of TDEs. Indeed, the replacement of $r_{\rm t} \rightarrow r_{\rm p}$ in Equation \eqref{Tfb} by, e.g., \citet{evans89, ulmer99, lodato09, strubbe09, lodato11}, with $r_{\rm p}$ the pericenter distance of the star (which could be much less than $r_{\rm t}$), results in a gross underestimate of $T_{\rm fb}$ and a corresponding overestimate of the luminosity \citep{guillochon13, stone13, norman21}. On the other hand, for stars with a large central density (e.g., those that are highly evolved), the core should be able to better withstand the tidal shear of the SMBH compared to the star on average, resulting in a smaller value for the tidal radius than Equation \eqref{rtdef}. Indeed, \citet{norman21} suggested that since Equation \eqref{rtdef} can be written as $r_{\rm t} \simeq (M_{\bullet}/\rho_{\star})^{1/3}$ with $\rho_{\star} = M_{\star}/(4\pi R_{\star}^3/3)$ the average density, the \emph{core disruption radius} at which the high-density core (and thus the entire star) is destroyed should be replaced with $r_{\rm t} \simeq (M_{\bullet}/\rho_{\rm c})^{1/3}$, with $\rho_{\rm c}$ the central stellar density. \cite{lawsmith19, ryu20} reached the same conclusion on empirical grounds through comparisons to simulations.

Additionally, the probability of a star being scattered onto an orbit about a SMBH with a pericenter distance $r_{\rm p} \equiv r_{\rm t}/\beta$ has a strong dependence on $\beta$: in the Newtonian approximation the probability distribution function of $\beta$ satisfies $f_{\beta} = \beta^{-2}$ for stars in the pinhole regime of scattering (e.g., \citealt{frank76, lightman77}), while relativistic effects cause $f_{\beta}$ to fall off even more steeply when the tidal radius is comparable to the direct capture radius of $4GM_{\bullet}/c^2$ for a non-spinning SMBH (which is particularly relevant for $M_{\bullet} \gtrsim 10^{7}M_{\odot}$; \citealt{coughlin22}). If the star is not destroyed at $r_{\rm t}$ but at $r_{\rm t}/\beta_{\rm c}$, then even if $\beta_{\rm c}$ is only marginally greater than 1 (see Table \ref{tab:1} below), a substantial fraction of TDEs will be partial and leave a stellar core intact. In these cases the rate at which stellar debris from the TDE is supplied to the SMBH, which should be comparable to the accretion luminosity, declines as $\propto t^{-9/4}$ \citep{coughlin19, miles20, nixon21}, which is significantly steeper than the canonical rate of $t^{-5/3}$ \citep{phinney89}. 

The precise value of $r_{\rm t}$ can thus have a large impact on the observable properties of TDEs. Here we argue that the distance at which the star is completely destroyed by tides can be more accurately (than Equation \ref{rtdef}) identified by equating the tidal field of the SMBH (including order-unity factors) to the maximum self-gravitational field within the star, which occurs at a distance within the stellar interior that we denote the core radius $R_{\rm c}$. This radius (and the maximum self-gravitational field) can be determined numerically and straightforwardly for any star, but, as we show below, is approximately given by $R_{\rm c} = R_{\star}\left(\rho_{\rm c}/\rho_{\star}\right)^{-1/3}$, and results in a ``core disruption radius'' that is approximately $r_{\rm t, c} \simeq \left(M_{\bullet}/\rho_{\rm c}\right)^{1/3}$, and a core disruption $\beta$ of $\beta_{\rm c} \simeq \left[\rho_{\rm c}/(4\rho_{\star})\right]^{1/3}$. In Section \ref{sec:analytic} we present our analysis, our results, and make comparisons to numerical simulations, and we summarize and conclude in Section \ref{sec:summary}.

\section{The core disruption radius and peak fallback properties}
\label{sec:analytic}
We define the tidal field as the difference in the gravitational field of the SMBH across the stellar diameter:

\begin{equation}
f_{\rm t} = \frac{GM_{\bullet}}{\left(r-R_{\star}\right)^2}-\frac{GM_{\bullet}}{\left(r+R_{\star}\right)^2} \simeq \frac{4GM_{\bullet}R_{\star}}{r^3}, \label{ftidal}
\end{equation}
where $M_{\bullet}$ is the SMBH mass, $r$ is the distance of the center of mass of the star to the SMBH, and $R_{\star}$ is the stellar radius, and in the last line we assumed $r \gg R_{\star}$. Typically the tidal field is defined as the difference in the gravitational field across the stellar radius, and the factor of 4 in Equation \eqref{ftidal} is usually a factor of 2. We argue that the factor of 4 treats the star as a material body and accounts for the fact that tides induce a velocity divergence across its diameter. As we also show below, this definition accurately reproduces the results of numerical, hydrodynamical simulations. 

\begin{figure} 
   \centering
   \includegraphics[width=0.475\textwidth]{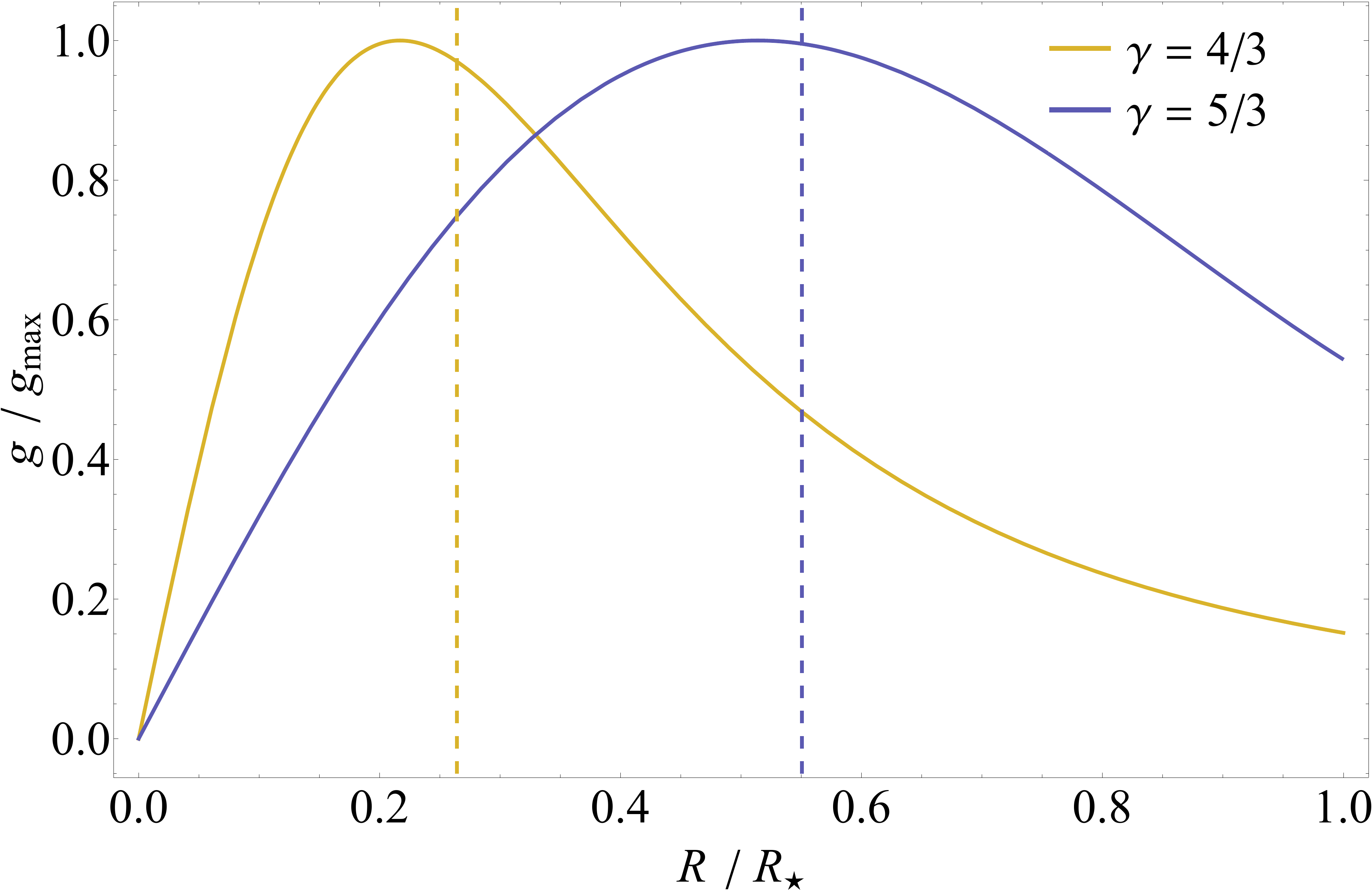} 
   \caption{The gravitational field within a $\gamma = 4/3$ (yellow) and $\gamma = 5/3$ (purple) polytrope normalized by its maximum value. The vertical, dashed lines give the approximate locations at which the gravitational field is maximized.}
   \label{fig:g_gmax}
\end{figure}

The canonical tidal radius equates the tidal field to the stellar surface gravity and drops numerical factors, yielding Equation \eqref{rtdef}. In general, however, a star's gravitational field is maximized in its interior, not at its surface. This is apparent from the fact that for radii within the star $R \simeq 0$, the gravitational field is $g(R) \simeq 4\pi G\rho_{\rm c}R/3$ with $\rho_{\rm c}$ the central stellar density, while for $R \lesssim R_{\star}$ we have $g(R) \simeq GM_{\star}/R^2$. Equating these two expressions for $g(R)$ then yields the approximate radius at which the gravitational field is maximized, which we define as the core radius, $R_{\rm c}$:

\begin{equation}
    R_{\rm c} \simeq 
     R_{\star}\left(\frac{\rho_{\rm c}}{\rho_{\star}}\right)^{-1/3}. \label{rmax}
\end{equation}
Here $\rho_{\star} = M_{\star}/(4\pi R_{\star}^3/3)$ is the average stellar density, and since $\rho_{\rm c} \ge \rho_{\star}$, we have $R_{\rm c} \le R_{\star}$. Figure \ref{fig:g_gmax} shows the gravitational field of $\gamma = 4/3$ and $5/3$ polytropes (so the stellar pressure $p$ and density $\rho$ are related via $p \propto \rho^{\gamma}$) normalized by their maximum values. The vertical, dashed lines show $R_{\rm c}$ as given by Equation \eqref{rmax}, and are $R_{\rm c}/R_{\star} \simeq 0.26$ and $R_{\rm c}/R_{\rm max} \simeq 0.55$ for the $\gamma = 4/3$ and $\gamma = 5/3$ polytrope, respectively, which slightly overestimate the true locations $R_{\rm c}/R_{\star} \simeq 0.22$ and $R_{\rm c}/R_{\star} \simeq 0.51$.

We expect the star to be completely destroyed when the tidal field evaluated at the core radius equals the self-gravity of the core, which corresponds to a tidal radius $r_{\rm t, c}$ of

\begin{equation}
    \frac{4GM_{\bullet}R_{\rm c}}{r_{\rm t, c}^3} \simeq \frac{4}{3}\pi G\rho_{\rm c}R_{\rm c} 
    \,\,\, \Rightarrow \,\,\, r_{\rm t, c} =  r_{\rm t}\left(\frac{\rho_{\rm c}}{4\rho_{\star}}\right)^{-1/3}. \label{rtcapp}
\end{equation}
Defining $\beta_{\rm c} = r_{\rm t}/r_{\rm t, c}$, we expect the core (and the entire star) to be destroyed when

\begin{equation}
    \beta_{\rm c, \,app} \simeq \left(\frac{\rho_{\rm c}}{4\rho_{\star}}\right)^{1/3}. \label{betacrit}
\end{equation}
\citet{lawsmith19} and \citet{ryu20} empirically found a similar form for the $\beta$ at which the star is completely disrupted by fitting the results of numerical simulations. Equation \eqref{betacrit} is approximate, as we extrapolated and equated the linear variation in the gravitational field from $R = 0$ to the inverse-square behavior from $R = R_{\star}$. More generally the core radius is where the self-gravitational field is maximized, and the core/complete disruption radius is found by equating the tidal and self-gravitational fields at that radius:

\begin{equation}
    \frac{4GM_{\bullet} R_{\rm c}}{r_{\rm t, c}^3} = g(R_{\rm c}) \quad \Rightarrow \quad 
    r_{\rm t, c} = r_{\rm t}\left(\frac{4GM_{\star}R_{\rm c}}{g(R_{\rm c})R_{\star}^3}\right)^{1/3},  \label{rtgen}
\end{equation}
corresponding to a $\beta$ of

\begin{equation}
\beta_{\rm c} = \frac{r_{\rm t}}{r_{\rm t, c}} = \left(\frac{4GM_{\star}R_{\rm c}}{g(R_{\rm c})R_{\star}^3}\right)^{-1/3}. \label{betac}
\end{equation}

\begin{figure*}
   \centering
   \includegraphics[width=0.495\textwidth]{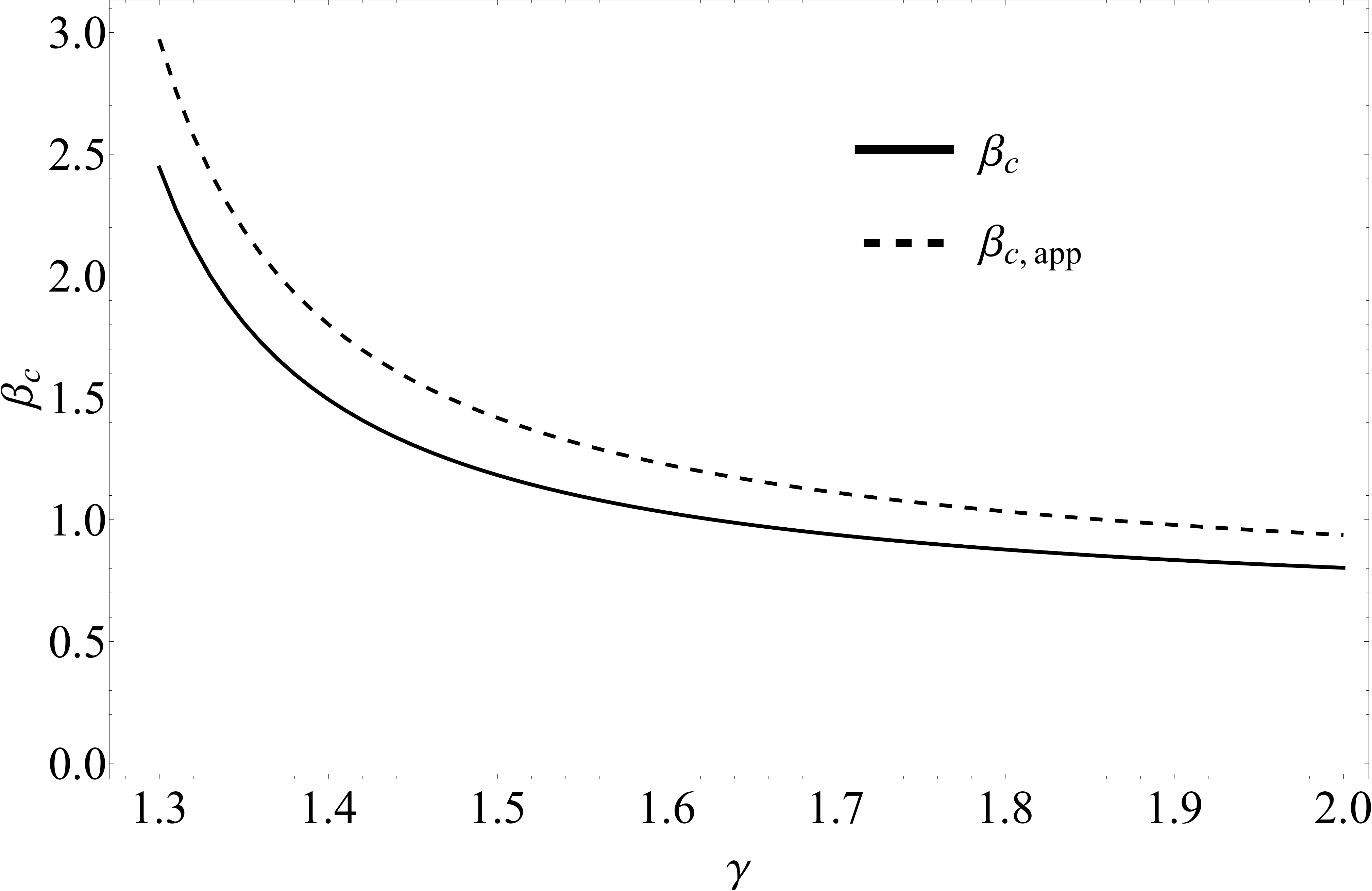} 
   \includegraphics[width=0.482\textwidth]{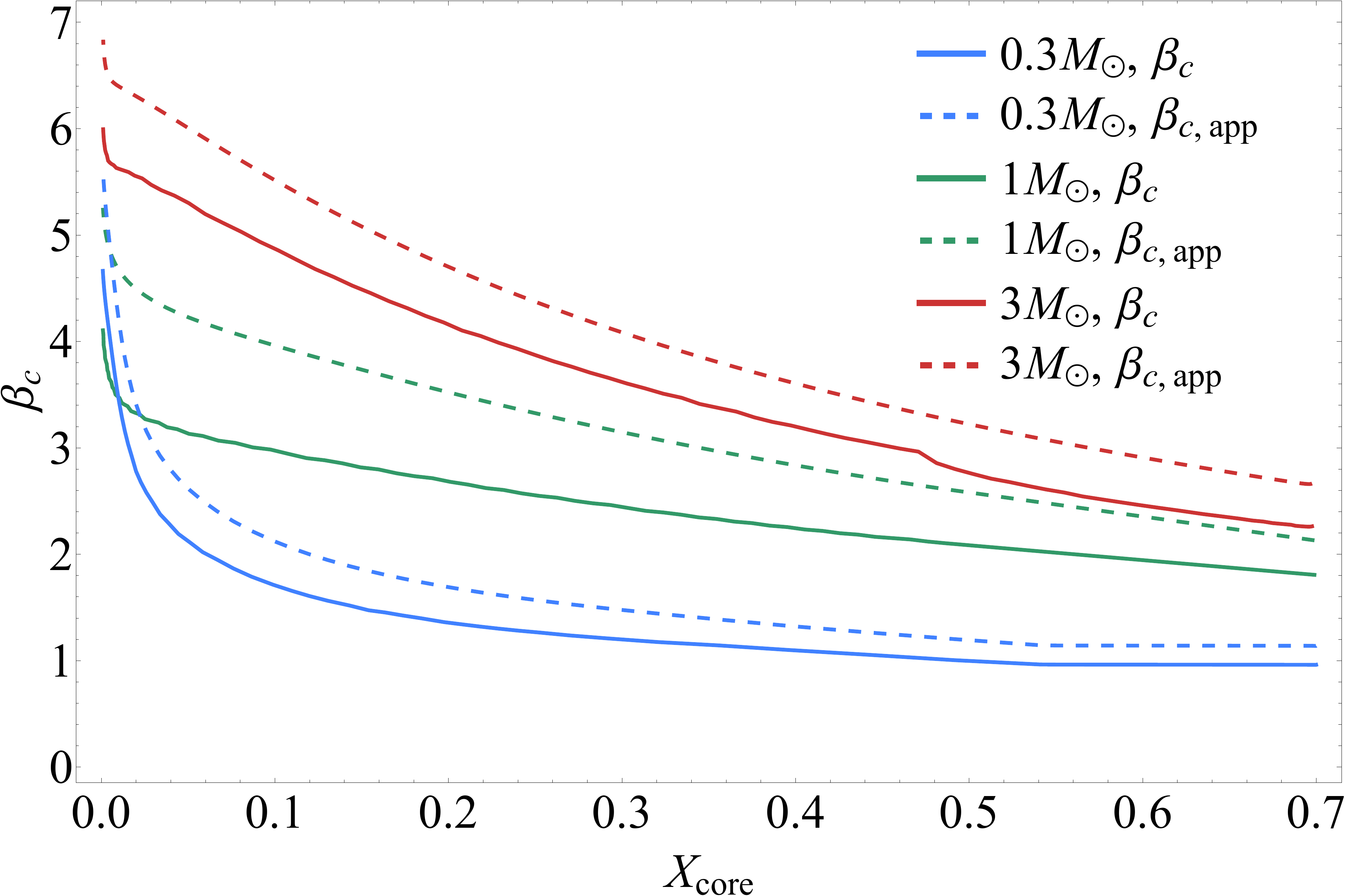} 
   \caption{Left: The $\beta$ at which a polytrope with polytropic index $\gamma$ is completely destroyed; the solid curve gives the exact expression, while the dashed curve is the approximate value that uses only the ratio of the central to the average density of the star (Equation \ref{betacrit}). Right: The core disruption $\beta$ as a function of the central Hydrogen mass fraction $X_{\rm core}$ of a $1M_{\odot}$ (green) and $0.3M_{\odot}$ (blue) star evolved through the main sequence with the stellar evolution code {\sc mesa}; each star begins on the main sequence at $X_{\rm core} \simeq 0.7$ and ends its main sequence evolution at $X_{\rm core} \simeq 0$. The solid curves are found by numerically calculating the maximum gravitational field within the star, while the dashed curves give the approximate solution obtained by using only the central density of the star. }
   \label{fig:betac}
\end{figure*}

The left panel of Figure \ref{fig:betac} shows $\beta_{\rm c}$ (Equation \ref{betac}, solid curve) and $\beta_{\rm c, app}$ (Equation \ref{betacrit}, dashed curve) for a polytrope with polytropic index $\gamma$. For a $\gamma = 5/3$ ($\gamma = 4/3$) polytrope, we have $\beta_{\rm c} \simeq 0.96$ ($\beta_{\rm c} \simeq 1.97$), while the approximate expression yields $\beta_{\rm c, app} \simeq 1.14$ ($\beta_{\rm c, app} \simeq 2.38$). By comparison, numerical hydrodynamical simulations find that the $\beta$ at which a $\gamma = 5/3$ polytrope is completely destroyed is $\beta_{\rm c, num} \simeq 0.92$ \citep{guillochon13, mainetti17, miles20}, while for a $\gamma = 4/3$ polytrope $\beta_{\rm c, num} \simeq 2$ \citep{guillochon13, mainetti17}. The right panel of Figure \ref{fig:betac} gives the exact and approximate $\beta_{\rm c}$ for a $0.3M_{\odot}$ (blue), $1M_{\odot}$ (green) and $3 M_{\odot}$ (red) star as a function of the Hydrogen mass fraction in its core $X_{\rm core}$, where each star was evolved in isolation at solar metallicity with the stellar evolution code {\sc mesa} \citep{paxton11} (v.~r21.12.1). The zero-age main sequence (ZAMS) corresponds to $X_{\rm core} \simeq 0.7$, while the terminal-age main sequence (TAMS) has $X_{\rm core} \simeq 0$. Over the lifetime of each star, $\beta_{\rm c}$ increases owing to the increasing density of the core, and does so dramatically near the TAMS. Table \ref{tab:1} gives the exact and approximate values of $\beta_{\rm c}$ for each star at ZAMS and TAMS and also at the ``middle-age main sequence'' (MAMS), defined to be where $X_{\rm core} \simeq 0.2$ (the $0.3M_{\odot}$, ZAMS star is effectively a $\gamma = 5/3$ polytrope and has the same $\beta_{\rm c}$ etc.~as the top row). The numerically obtained values for the $1M_{\odot}$ ZAMS, $1M_{\odot}$ MAMS, and $0.3M_{\odot}$ MAMS are taken from \citet{nixon21}, while upper limits are from \citet{golightly19}. 

\begin{table}
\centering
\begin{tabular}{|c|c|c|c|c|c|}
\hline
\textrm{star} & $\beta_{\rm c} $& $\beta_{\rm c, app}$ & $\beta_{\rm c, num}$ & $T_{\rm peak}$ & $\dot{M}_{\rm peak}$\\ 
\hline
5/3 polytrope & 0.96 & 1.14 & 0.92 & 62 d & 1.5 $M_{\odot}$ yr$^{-1}$\\ 
\hline
4/3 polytrope & 1.97 & 2.38 & 2 & 27 & 3.4 \\ 
\hline
$0.3M_{\odot}$ MAMS & 1.34 & 1.67 & 1.6 & 36 & 0.76 \\ 
\hline
$0.3M_{\odot}$ TAMS & 4.7 & 5.6 & $>3$ & 15 & 1.8 \\
\hline
$1M_{\odot}$ ZAMS & 1.80 & 2.13 & 1.79 & 24 & 3.8 \\ 
\hline
$1M_{\odot}$ MAMS & 2.7 & 3.5 & 3.5 & 23 & 4.0 \\ 
\hline
$1M_{\odot}$ TAMS & 4.1 & 5.2 & $> 3$ & 25 & 3.8  \\ 
\hline
$3M_{\odot}$ ZAMS & 2.26 & 2.66 & $< 3$ & 18 & 15 \\ 
\hline
$3M_{\odot}$ MAMS & 4.1 & 4.6 & $>3$ & 27 & 10 \\ 
\hline
$3M_{\odot}$ TAMS & 6 & 6.8 & $> 3$ & 21 & 13  \\ 
\hline
\end{tabular}
\caption{The predicted $\beta$ at which the core of the star is destroyed $\beta_{\rm c}$, the approximate value at which it is destroyed $\beta_{\rm c, app}$, and the value of $\beta$ at which the star is destroyed as obtained from numerical, hydrodynamical simulations, $\beta_{\rm c, num}$, for the type of star shown in the left column. $T_{\rm peak}$ and $\dot{M}_{\rm peak}$ give the time to the peak fallback rate and its value, calculated with Equation \eqref{Tpeak}, when the star is disrupted by a $10^6M_{\odot}$ SMBH. }
\label{tab:1}
\end{table}

\begin{figure*} 
   \centering
   \includegraphics[width=0.49\textwidth]{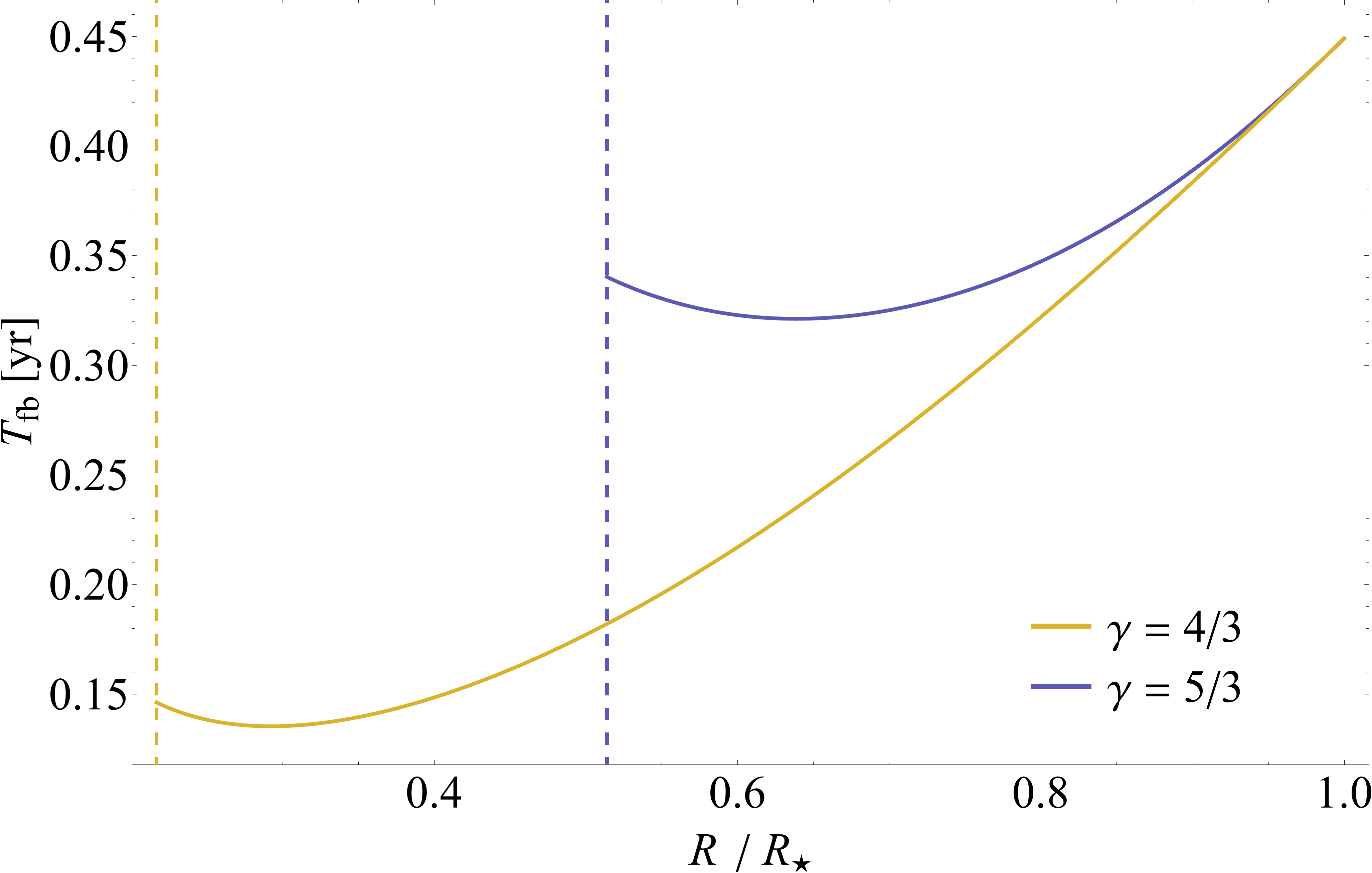} 
   \includegraphics[width=0.485\textwidth]{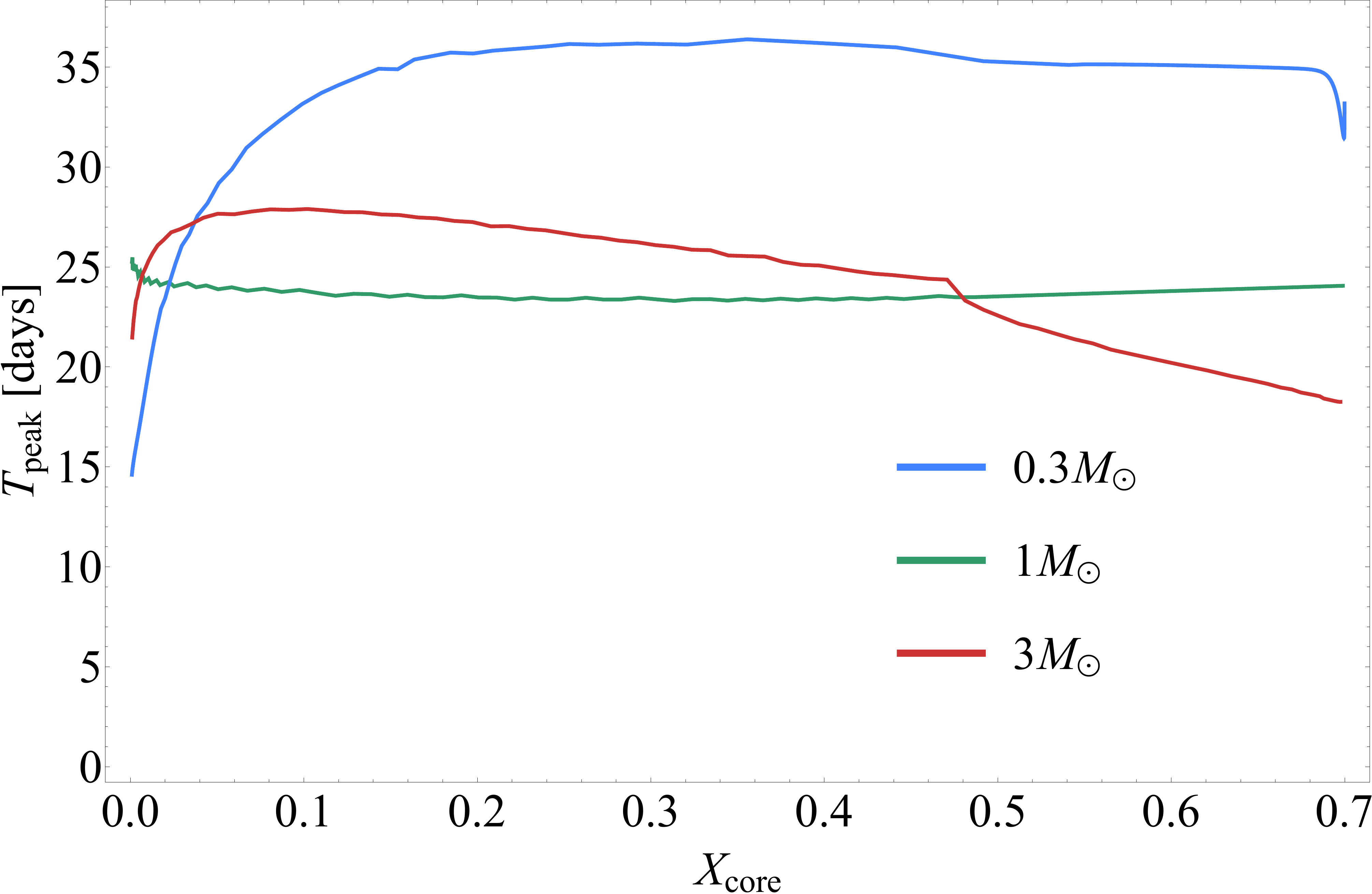} 
   \caption{Left: The fallback time as a function of initial radius within the star for a $\gamma = 4/3$ (yellow) and $\gamma = 5/3$ (blue) solar-like polytrope disrupted by a $10^{6}M_{\odot}$ SMBH. The vertical, dashed lines give the location of the core. Right: The peak fallback time as a function of the Hydrogen core fraction for the same three stars as in the right panel of Figure \ref{fig:betac}.}
   \label{fig:TfbofR}
\end{figure*}

The fallback time given in Equation \eqref{Tfb} is estimated by making the (crude; see \citealt{steinberg19}) approximation that upon passing through $r_{\rm t}$ the entire star moves with the center of mass and is undistorted, in which case the energy of each fluid element is ``frozen-in'' at the tidal radius and calculable as a function of its position within the star. Here, however, when the center of mass reaches $r_{\rm t, c}$ we do not expect this model to be even approximately correct for the layers of the star that are outside of the core radius, as these fluid shells have already been overcome by the gravitational field of the SMBH. We can gain some insight into the complexity that this aspect adds to the problem by assuming that the energy of each fluid shell at radii $R > R_{\rm c}$ is established at its tidal radius, i.e., that the tidal radius as a function of spherical $R$ (valid for $R > R_{\rm c}$), and the corresponding fallback time, is (from Equations \ref{rtgen} and \ref{Tfb})

\begin{equation}
r_{\rm t}(R) = \left(\frac{4GM_{\bullet}R}{g(R)}\right)^{1/3}, \quad T_{\rm fb}(R) = \left(\frac{r_{\rm t}(R)^2}{2R}\right)^{3/2}\frac{2\pi} {\sqrt{GM_{\bullet}}}. \label{rtofR}
\end{equation}
The left panel of Figure \ref{fig:TfbofR} shows $T_{\rm fb}(R)$ for a $\gamma = 4/3$ (yellow) and $5/3$ (purple) polytrope that has $R_{\star} = R_{\odot}$, $M_{\star} = M_{\odot}$, and $M_{\bullet} = 10^{6}M_{\odot}$, and the vertical, dashed lines show the location of the core radius. We see that the fallback time decreases from the surface and reaches a relative minimum at a location near, but just outside of, the core radius. However, this model for the outer layers cannot possibly be correct, because the extremities of the star are closer to the SMBH at the time of disruption. If the fallback times were distributed as suggested by the left panel of Figure \ref{fig:TfbofR}, fluid shells at smaller radii in the interior of the star would cross those at larger radii, which physically cannot happen. 

Figure \ref{fig:TfbofR} suggests that gas at radii $R \gtrsim R_{\rm c}$ must return to the SMBH on a timescale that is shorter than the minimum timescale reached by $T_{\rm fb}(R)$, but that the energy is distributed dynamically and in a way that is not captured with this model. The smallest possible value we would expect for the return time, $T_{\rm ret}$, is obtained by letting $R = R_{\star}$ and $r_{\rm t}(R) = r_{\rm t, c}\beta_{\rm c}$ in Equation \eqref{rtofR}, i.e., 

\begin{equation}
T_{\rm ret} \gtrsim T_{\rm fb}(R_{\star})\beta_{\rm c}^{-3}. \label{Tret}
\end{equation}
Note that this is the same value one would obtain by assuming that the energy is frozen-in at pericenter. However, we are not arguing that this expression holds for any value of $\beta$; rather, it is the shortest return time we expect for the material provided that the center of mass reaches a pericenter distance smaller than $r_{\rm t, c}$. 

While the gas at $R \gtrsim R_{\rm c}$ likely evolves in a way that is not able to be accurately captured with this model (and Equation \ref{Tret} should be interpreted as a rough lower bound), the core (gas shells at $R\lesssim R_{\rm c}$) can still be approximated as moving with the center of mass until reaching $r_{\rm t, c}$ and should return to the SMBH on a timescale of $\sim T_{\rm fb}(R_{\rm c})$. Since the core contains a substantial fraction of the mass of the star (indeed, assuming a constant density $\rho_{\rm c}$ for $R \le R_{\rm c}$ gives $M_{\rm c} \simeq M_{\star}$), and hence a substantial fraction will have already accreted by that time, we expect $T_{\rm fb}(R_{\rm c})$ to coincide approximately with the \emph{peak} fallback time, or $T_{\rm fb}(R_{\rm c}) = NT_{\rm peak}$, where $N \sim 1$ is a constant numerical factor across all stars and determinable from hydrodynamical calculations. Remarkably, comparing $T_{\rm fb}(R_{\rm c})$ to $T_{\rm peak}$ from simulations in \citet{guillochon13}, \citet{coughlin15}, \citet{golightly19} and \citet{nixon21}, we find that $N = 2$ nearly exactly reproduces the numerically obtained peak fallback times for every star, and thus

\begin{equation}
T_{\rm peak} = \left(\frac{r_{\rm t}(R_{\rm c})^2}{2R_{\rm c}}\right)^{3/2}\frac{\pi} {\sqrt{GM_{\bullet}}}. \label{Tpeak}
\end{equation}

The right panel of Figure \ref{fig:TfbofR} shows the peak fallback time given by Equation \eqref{Tpeak} for a $10^6M_{\odot}$ SMBH and the same stars as in the right panel of Figure \ref{fig:betac} as a function of their core Hydrogen mass fraction; the values at ZAMS, MAMS, and TAMS are given in Table \ref{tab:1}. The striking feature of these curves is that they display much less variation with respect to $X_{\rm core}$ than does $\beta_{\rm c}$ (see the right panel of Figure \ref{fig:betac}), and the $1M_{\odot}$ star in particular has an almost constant peak fallback time at $\sim 24$ days. This finding is consistent with \citet{nixon21}, as the solid-blue and dashed-green curves in the middle panel of their Figure 3 are effectively identical for all $\beta \gtrsim 2$ (and equal to $\sim 25$ days; note that the legend for this figure is incorrect -- the dashed-green curve is for the $1M_{\odot}$ MAMS star). Equation \eqref{Tpeak} can also be substantially shorter than the peak fallback time derived from the frozen-in approximation, e.g., $T_{\rm peak} \simeq 24$ days for a $1M_{\odot}$ ZAMS star, whereas employing the frozen-in approximation yields $T_{\rm peak} \simeq 1$ year, which is over an order of magnitude longer; see Figure 2 of \citet{golightly19}.

We can also estimate the magnitude of the peak fallback rate: since half of the stellar mass is accreted during a TDE and roughly half of that mass will have been accreted by $T_{\rm peak}$, we expect

\begin{equation}
\dot{M}_{\rm peak} \simeq \frac{M_{\star}}{4T_{\rm fb}}. \label{Mdotpeak}
\end{equation}
The final column in Table \ref{tab:1} gives the peak fallback rate for each star; comparing to \citet{guillochon13, golightly19, nixon21} shows that these predictions are in remarkably good agreement with the results of numerical simulations. We also note that while the value of $\beta_{\rm c}$ in Table \ref{tab:1} is somewhat smaller than $\beta_{\rm num}$ for the $0.3M_{\odot}$ MAMS and $1.0M_{\odot}$ MAMS stars, the top panel of Figure 3 in \citet{nixon21} shows that $\beta_{\rm c}$ coincides almost exactly with the $\beta$ at which the fallback rate reaches its maximum value, which suggests that in these instances the core is largely destroyed and/or reforms at a later time and does not substantially affect the fallback. \citet{lawsmith19} also noted that very compact stars did not satisfy $\beta_{\rm c} \propto \left(\rho_{\rm c}/\rho_{\star}\right)^{1/3}$. 

The expression for $\beta_{\rm c}$ \eqref{betac} is only a function of the properties of the star. Therefore, $r_{\rm t, c}$, $T_{\rm peak}$, and $\dot{M}_{\rm peak}$ are valid for any SMBH mass, and this will only break down once the tidal radius becomes either comparable to the size of the star (i.e., the tidal approximation becomes invalid) or highly relativistic and the gravitational radius introduces an additional scale length. These two regimes are approached in the small and large-SMBH-mass limits, respectively.

Finally, our inferred distance at which the tidal field equals the self-gravitational field of the star at the stellar surface is a factor of $4^{1/3}$ larger than the canonical estimate. However, given our arguments, we would expect this distance to be the one at which the star just begins to lose mass. Therefore, the \emph{partial disruption radius}, where we expect any mass to be stripped from the envelope, is

\begin{equation}
\beta_{\rm partial} = 4^{-1/3} \simeq 0.6,
\end{equation}
independent of the stellar properties. This agrees with simulations, which find that the $\beta$ at which any mass loss occurs is $\beta \simeq 0.55-0.6$ (e.g., \citealt{guillochon13, nixon21}).

\section{Summary and Conclusions}
\label{sec:summary}
We proposed that the complete tidal disruption radius of a star can be accurately constrained by equating the SMBH tidal field (including a factor of 4 that accounts for the differential stretching across the stellar diameter) to the maximum self-gravitational field within the star, which is generally in the stellar interior. To our knowledge this statement has not been made in the literature. The radius at which this equality occurs, which we define as the core radius $R_{\rm c}$, can be straightforwardly determined numerically for any progenitor and its value (and the self-gravitational field at $R_{\rm c}$) inserted into Equation \eqref{betac} to determine $\beta_{\rm c}$, where $r_{\rm t}/\beta_{\rm c}$ -- with $r_{\rm t}$ the canonical tidal radius -- is the distance within which the star must come to be completely destroyed. We performed this exercise for a range of stellar progenitors, and we also calculated the peak fallback time and the magnitude of the peak fallback from the TDE (see Equations \ref{Tpeak} and \ref{Mdotpeak}) and found very good agreement with the results of hydrodynamical simulations, e.g., $\beta_{\rm c}\simeq 0.96$ ($\simeq 1.97$) for a $\gamma = 5/3$ ($4/3$) progenitor, while simulations yield $\beta_{\rm c, num} \simeq 0.92$ ($\beta_{\rm c, num} \simeq 2$). In general $\beta_{\rm c}$ must be calculated numerically as a function of the progenitor (and only of the progenitor, i.e., the SMBH mass does not enter, unless the SMBH mass is very small so that the tidal approximation breaks down, or very large so that relativistic effects become important), but it is approximately given by $\beta_{\rm c} \simeq \left[\rho_{\rm c}/(4\rho_{\star})\right]^{1/3}$, where $\rho_{\rm c}$ ($\rho_{\star}$) is the central (average) stellar density. 

For any stellar population, a scattering rate of stars into the loss cone of the SMBH, and the probability distribution function of the pericenter distance of tidally disrupted stars, the number of full vs.~partial disruptions can be determined via Equation \eqref{betac}. The relativistic distribution of pericenter distances was calculated by \citet{coughlin22} in the full loss cone regime and shown to drop sharply near the direct capture radius of the SMBH, and full disruptions are replaced by direct captures (i.e., the star is swallowed whole). Since high-$\beta$'s are required to disrupt high-mass ($M_{\star} \gtrsim 1M_{\odot}$) stars and the tidal radius is proportional to the stellar radius, which is smaller (and hence more relativistic) for low-mass stars, Figure \ref{fig:betac} suggests that the vast majority of disruptions by high-mass SMBHs will be partial and yield a fallback rate that scales as $\propto t^{-9/4}$.

With Equations \eqref{Tpeak} and \eqref{Mdotpeak} for the time and magnitude of the peak fallback rate, one can -- for a given observational facility and observing strategy -- estimate the number of \emph{observable} TDEs for a given, underlying SMBH mass distribution. We can also estimate the number of TDEs that will undergo a period of substantial super-Eddington accretion, and thus are likely to give rise to relativistic and jetted outflows. Such information is therefore extremely useful for constraining the demographics of SMBHs throughout cosmic time with high-cadence surveys such as the Rubin Observatory.

\section*{Data Availability}
The data underlying this article will be shared on reasonable request. 

\section*{Acknowledgements}
We thank DJ Pasham for useful comments and the referee for useful correspondence. E.R.C.~acknowledges support from the National Science Foundation through grant AST-2006684 and the Oakridge Associated Universities through a Ralph E.~Powe Junior Faculty Enhancement Award. C.J.N.~acknowledges support from the Science and Technology Facilities Council [grant number ST/W000857/1].

\bibliographystyle{mnras}

\bsp	% typesetting comment
\label{lastpage}
\end{document}